\documentclass[aps,english,preprint,nofootinbib,preprintnumbers]{revtex4}
\usepackage{mathrsfs}
\usepackage{hyperref}
\usepackage{graphicx}
\usepackage{amsmath, amssymb}
\usepackage{babel}
\usepackage{color}
\usepackage{slashed}

\def\beqn{\begin{eqnarray}}
\def\eeqn{\end{eqnarray}}
\def\barr{\begin{array}}
\def\earr{\end{array}}
\def\btab{\begin{tabular}}
\def\etab{\end{tabular}}
\def\bite{\begin{itemize}}
\def\eite{\end{itemize}}
\def\bcen{\begin{center}}
\def\ecen{\end{center}}

\begin{document}

\title{
Update on {\boldmath$|V_{us}|$} and {\boldmath$|V_{us}/V_{ud}|$} from semileptonic\\ kaon and pion decays}

\author{Chien-Yeah Seng$^{1}$}
\author{Daniel Galviz$^{1}$}
\author{William J. Marciano$^{2}$}
\author{Ulf-G. Mei{\ss}ner$^{1,3,4}$}

\affiliation{$^{1}$Helmholtz-Institut f\"{u}r Strahlen- und Kernphysik and Bethe Center for
  Theoretical Physics,\\
	Universit\"{a}t Bonn, 53115 Bonn, Germany}
\affiliation{$^{2}$Department of Physics, Brookhaven National Laboratory, Upton, New York 11973, USA}
\affiliation{$^{3}$Institute for Advanced Simulation, Institut f\"ur Kernphysik and J\"ulich
  Center for Hadron Physics,
 Forschungszentrum J\"ulich, 52425 J\"ulich, Germany}
\affiliation{$^{4}$Tbilisi State  University,  0186 Tbilisi, Georgia}

\date{\today}

\begin{abstract}

Implications for Cabibbo universality based on progress in the study of semileptonic kaon and pion decays are discussed. Included are recent updates of experimental input along with improved  radiative corrections, form factors and isospin breaking effects.  As a result, we obtain for the
Cabibbo-Kobayashi-Maskawa (CKM) quark mixing matrix element $|V_{us}| = 0.22309(56)$ from semileptonic $K_{\ell 3}$ $(K\rightarrow \pi\ell\nu)$ decays and $|V_{us}/V_{ud}| = 0.22908(87)$ from the ratio between the kaon and pion ($\pi_{e3}$) semileptonic decay rates. In both, a lattice QCD value of the
form factor $|f_+^K(0)|=0.9698(17)$ is employed. The $V_{us}$ from $K_{\ell 3}$ decays together with $V_{ud}=0.97373(31)$ found from superallowed nuclear beta decays implies an apparent $3.2\sigma$ violation of the first-row CKM unitarity condition. The $|V_{us}|/|V_{ud}|$ obtained from the ratio of weak vector current induced meson decays is consistent with the observed unitarity violation but found to differ by $2.2\sigma$ from its extraction using the ratio of weak axial-vector leptonic decay rates $\Gamma(K\rightarrow \mu\nu)/\Gamma(\pi\rightarrow\mu\nu)$.  The situation suggests a difference between vector and axial-vector derived CKM matrix elements or a problem with the lattice QCD form factor input. Prospects for future improvements in comparative precision tests involving $|V_{ud}|$, $|V_{us}|$ and their ratios are briefly described.

\end{abstract}

\maketitle


\section{Introduction}

Precision tests of the Standard Model (SM) have become increasingly important given the null results
from high-energy colliders in the search for physics beyond the Standard Model (BSM)~\cite{Zyla:2020zbs}.
Deviation from expectations in the muon anomalous magnetic moment~\cite{Fermigm2,Aoyama:2020ynm,Miller:2007kk,Miller:2012opa,Jegerlehner:2009ry}, hints of lepton flavor universality violation in
B decays~\cite{Aaij:2019wad,Aaij:2014ora,Aaij:2015yra,Aaij:2015oid} and tests of unitarity in the
Cabibbo-Kobayashi-Maskawa (CKM) quark mixing matrix are all exhibiting potential BSM effects.
In the last case, improvements in the electroweak radiative corrections (RCs)~\cite{Seng:2018yzq,Seng:2018qru,Czarnecki:2019mwq,Seng:2020wjq,Hayen:2020cxh,Hayen:2021iga,Shiells:2020fqp,Seng:2021syx} have revealed tension in the first-row
unitarity requirement $|V_{ud}|^2 + |V_{us}|^2 +|V_{ub}|^2 =1$~\cite{Hardy:2020qwl}. Similarly, the difference in the
value of $|V_{us}|$ extracted from $K_{\ell 3}$ and $K_{\mu2}$ decays needs to be better understood\footnote{All decay processes are understood to be radiative inclusive; for example, $K_{\ell 3}$ means $K\rightarrow \pi\ell\nu(\gamma)$.}.  Possible
explanations based on BSM calculations~\cite{Bryman:2019ssi,Bryman:2019bjg,Kirk:2020wdk,Grossman:2019bzp,Belfatto:2019swo,Cheung:2020vqm,Jho:2020jsa,Yue:2020wkj,Endo:2020tkb,Capdevila:2020rrl,Eberhardt:2020dat,Crivellin:2020lzu,Coutinho:2019aiy,Gonzalez-Alonso:2018omy,Falkowski:2019xoe,Cirgiliano:2019nyn,Falkowski:2020pma,Becirevic:2020rzi,Crivellin:2021njn,Tan:2019yqp,Crivellin:2021bkd,Crivellin:2020ebi,Crivellin:2020klg,Dekens:2021bro,Belfatto:2021jhf} have been conjectured. The confirmation of such ideas will require improvements in both experiment and SM theory. 

In this work, we focus on the extraction of $|V_{us}|$ from $K\rightarrow\pi\ell\nu$ semileptonic
decay processes known as $K_{\ell 3}$. Our starting point is the comprehensive 2010 FlaviaNet Working Group Report~\cite{Antonelli:2010yf} which presented a thorough review of all relevant experimental and theoretical information available at the time. Since then, significant progress has been made on both the experimental and theoretical fronts, including new measurements
of the $K_S$ lifetime~\cite{KTeV:2010sng,KLOE:2010yit} and branching ratio (BR)~\cite{KLOE-2:2019rev}, updates of the electroweak RCs~\cite{Ma:2021azh,Seng:2021boy,Seng:2021wcf},
phase space factors~\cite{NewPS},  $K\pi$ form factor and isospin-breaking corrections
(\cite{FlavourLatticeAveragingGroup:2019iem} and references therein). Some of those advances have been more recently discussed by FlaviaNet Working Group members in the form of proceedings ~\cite{Moulson:2017ive} and conference slides~\cite{CKM18,NewPS} rather than more detailed research publications, making cross-checking difficult. 
For that reason, we present in this paper an updated status report on the value of $|V_{us}|$ extracted from $K_{\ell 3}$ decay properties.
With the recent theory progress, we find that apart from the lattice input of the $K\pi$ form factor
which is a universal multiplicative constant, the experimental errors from the kaon lifetimes
and BRs are by far the dominant sources of uncertainties in \textit{all} six channels of
$K_{\ell 3}$, which is quite different from the situation a few years ago, in particular before the new calculation of the $K_{e3}$ RC~\cite{Seng:2021boy,Seng:2021wcf}.

The progress described above can also be applied to the determination of the recently proposed ratio
$R_V=\Gamma(K_{\ell 3})/\Gamma(\pi_{e3})$ as an alternative approach to study $|V_{us}/V_{ud}|$~\cite{Czarnecki:2019iwz},
complementary to the existing method based on $R_A=\Gamma(K_{\mu 2})/\Gamma(\pi_{\mu 2})$. We show,
following the recent improvements in the precision level of the
$K_{e3}$~\cite{Seng:2021boy,Seng:2021wcf} and $\pi_{e3}$ RC~\cite{Feng:2020zdc}, that $R_V$ is
now a theoretically cleaner observable than $R_A$. It provides strong motivation to further improve the experimental precision of the $\pi_{e3}$ BR and $K_{\ell 3}$ decay properties as much as possible. An experimental next-generation rare pion decay program proposal: PIONEER~\cite{Pioneer,ArevaloSnowmass} would aim to improve the experimental $\pi_{e3}$ decay rate by a factor of 3 or better, making the $K_{\ell 3}$ decay rate the dominant uncertainty in $R_V$.

The paper is organized as follows: In Sec.~\ref{sec:Vus}, we give a detailed update on 
 $|V_{us}f_{+}^{K}(0)|$ and $|V_{us}|$. Similarly, Sec.~\ref{sec:Vus/Vud} provides
updates on $|V_{us}f_{+}^{K}(0)|/|V_{ud}f_{+}^{\pi}(0)|$ and $|V_{us}/V_{ud}|$. We end with 
conclusions in Sec.~\ref{sec:con}. Some technicalities are relegated to the Appendix.

\section{Updates on $|V_{us}f_{+}^{K}(0)|$ and $|V_{us}|$}
\label{sec:Vus}

We start from the quantity $|V_{us}f_{+}^{K}(0)|$, with $f_+^{K}(0)$ the vector form factor in
the $K^0\to \pi^-$ transition at zero momentum transfer. It is extracted from $K_{\ell3}$ decays
through the following master formula~\cite{Zyla:2020zbs}:
\begin{equation}
A_{K\ell}\equiv|V_{us}f_{+}^{K}(0)|_{K\ell}=\sqrt{\frac{192\pi^{3}\mathrm{BR}(K\ell)\Gamma_K}{G_{F}^{2}M_{K}^{5}C_{K}^{2}S_{\mathrm{EW}}I_{K\ell}(1+\delta_{\mathrm{EM}}^{K\ell}+\delta_{\mathrm{SU(2)}}^{K\ell})}}~,\label{eq:Kl3master}
\end{equation}
where $G_F=1.1663787(6)\times 10^{-5}$~GeV$^{-2}$ is the Fermi constant obtained from muon decay,
$\Gamma_K$ the total kaon decay width, BR($K\ell$) the $K_{\ell 3}$ branching ratio,
and $C_K$ a simple isospin factor which equals 1 ($1/\sqrt{2}$) for $K^0$ ($K^+$) decay. The SM theory inputs to the right-hand side are as follows:
$S_{\mathrm{EW}}=1.0232(3)_{\mathrm{HO}}$ is a universal short-distance
EW factor~\cite{Marciano:1993sh}, and the uncertainty comes from higher-order QED
effects~\cite{Erler:2002mv} which is common to all channels and will not take part in
the weighted average. Meanwhile, $I_{K\ell}$, $\delta_\mathrm{EM}^{K\ell}$ and
$\delta_\mathrm{SU(2)}^{K\ell}$ are the phase space factor, the long-distance electromagnetic (EM)
correction and the isospin-breaking correction, respectively. These are channel-specific inputs.
To facilitate the discussion of correlation effects, we group the values of $A_{K\ell}$ from
six independent $K_{\ell 3}$ channels into a vector:
\begin{equation}
A=(A_{K_{L}e},A_{K_{S}e},A_{K^{+}e},A_{K_{L}\mu},A_{K_{S}\mu},A_{K^{+}\mu})^{T}~.
\end{equation}
The order of the entries is important, as is seen later. 

In what follows, we summarize all the data input needed in this work. All the experimental
data of decay widths and BRs are obtained from the 2021 online update~\cite{PDGonline} of
the Particle Data Group (PDG) review~\cite{Zyla:2020zbs}. Knowing that different choices of inputs of
statistical analyses of the same problem may lead to different quantitative conclusions, throughout the discussion we will explain the similarities and differences in  the data inputs between our work and existing global analysis (in particular the 2010 FlaviaNet review~\cite{Antonelli:2010yf} and its updates~\cite{Moulson:2017ive,CKM18,NewPS}), and 
present all the essential steps in some detail despite that most of them are familiar to
experts; the basic mathematical tools needed in this
work are summarized in the Appendix.  With such, all the intermediate and final results in this paper are fully transparent and can be easily crossed-checked by interested readers, or compared to similar analyses with
possibly different statistical approaches. 

\subsection{{\boldmath$K_{L}$} experimental inputs}

The $K_L$ decay width quoted in PDG is obtained through a combined fit from
Refs.~\cite{KLOE:2005vdt,KLOE:2005lau,Vosburgh:1972zqy},
whereas the $K_Le$, $K_L\mu$ BRs are fitted from Refs.~\cite{KLOE:2005vdt,KTeV:2004hpx}, all which have been included in the FlaviaNet 2010 review.
The results read:
\begin{eqnarray}
	K_{L}^{\mathrm{exp}} & = & \left(\begin{array}{ccc}
		\mathrm{BR}(K_{L}e) & \mathrm{BR}(K_{L}\mu) & \Gamma_{K_{L}}\end{array}\right)^{T}\nonumber\\
	& = & \left(\begin{array}{ccc}
		0.4055(11) & 0.2704(7) & 1.2866(53)\times 10^{-14}\:\mathrm{MeV}\end{array}\right)^{T}
\end{eqnarray}
with the correlation matrix (it is symmetric, so we only show the upper right components for simplicity)
\begin{equation}
\mathrm{Corr}(K_{L}^{\mathrm{exp}})=\left(\begin{array}{ccc}
1 & -0.2149932 & -0.2676291\\
& 1 & -0.08759115\\
&  & 1
\end{array}\right)
\end{equation}
from which we can obtain the covariance matrix $\mathrm{Cov}(K_{L}^{\mathrm{exp}})$
using Eq.\ref{eq:CorrtoCov}. The
contribution of $K_{L}^{\mathrm{exp}}$ to the covariance matrix of
$A$ is given by:
\begin{equation}
\mathrm{Cov}(A)_{K_{L}^{\mathrm{exp}}}=\left(\frac{\partial A}{\partial K_{L}^{\mathrm{exp}}}\right)\cdot\mathrm{Cov}(K_{L}^{\mathrm{exp}})\cdot\left(\frac{\partial A}{\partial K_{L}^{\mathrm{exp}}}\right)^{T}
\end{equation}
where 
\begin{equation}
\left(\frac{\partial A}{\partial K_{L}^{\mathrm{exp}}}\right)=\left(\begin{array}{ccc}
\frac{\partial A_{K_{L}e}}{\partial\mathrm{BR}(K_{L}e)} & 0 & \frac{\partial A_{K_{L}e}}{\partial\Gamma_{K_{L}}}\\
0 & 0 & 0\\
0 & 0 & 0\\
0 & \frac{\partial A_{K_{L}\mu}}{\partial\mathrm{BR}(K_{L}\mu)} & \frac{\partial A_{K_{L}\mu}}{\partial\Gamma_{K_{L}}}\\
0 & 0 & 0\\
0 & 0 & 0
\end{array}\right)~.
\end{equation}

\subsection{{\boldmath$K_{S}$} experimental inputs}

The $K_S$ decay width quoted in PDG is fitted from
Refs.~\cite{KTeV:2010sng,KLOE:2010yit,NA48:2002iol,Bertanza:1996dt,Schwingenheuer:1995uf,Gibbons:1993zj},
and the $K_Se$ BR from Refs.~\cite{Batley:2007zzb,KLOE:2006vvm,KLOE:2002lao}. Notice that not all of them were utilized in the FlaviaNet review and updates, but only the more recent results from the KTeV~\cite{KTeV:2010sng}, KLOE~\cite{KLOE:2006vvm,KLOE:2010yit} and NA48~\cite{NA48:2002iol,Batley:2007zzb} collaborations. Also, only five channels in $K_{\ell 3}$ were analyzed in all the past
reviews because the $K_S\mu$ BR was not
independently measured. The first direct measurement of this BR appeared in year 2020~\cite{KLOE-2:2019rev}, which allows us to finally include all six channels in the combined
analysis of $V_{us}$ for the first time. The experimental results read:
\begin{eqnarray}
K_{S}^{\mathrm{exp}} & = & \left(\begin{array}{ccc}
\mathrm{BR}(K_{S}e) & \mathrm{BR}(K_{S}\mu) & \Gamma_{K_{S}}\end{array}\right)^{T}\nonumber\\
& = & \left(\begin{array}{ccc}
7.04(8)\times 10^{-4} & 4.56(20)\times 10^{-4} & 7.3510(33)\times 10^{-12}\:\mathrm{MeV}\end{array}\right)^{T}~.
\end{eqnarray}
The PDG provides the correlation coefficient between the two BRs, but not between the
BRs and the total decay width. The latter is expected to be very small from
Ref.~\cite{Antonelli:2010yf}, so here we simply set it to zero. With that, we obtain the
following correlation matrix:
\begin{equation}
\mathrm{Corr}(K_{S}^{\mathrm{exp}})=\left(\begin{array}{ccc}
1 & -0.00144257 & 0\\
& 1 & 0\\
& & 1
\end{array}\right)~.
\end{equation}
The contribution of $K_{S}^{\mathrm{exp}}$ to the covariance matrix
of $A$ is given by:
\begin{equation}
\mathrm{Cov}(A)_{K_{S}^{\mathrm{exp}}}=\left(\frac{\partial A}{\partial K_{S}^{\mathrm{exp}}}\right)\cdot\mathrm{Cov}(K_{S}^{\mathrm{exp}})\cdot\left(\frac{\partial A}{\partial K_{S}^{\mathrm{exp}}}\right)^{T}
\end{equation}
where 
\begin{equation}
\left(\frac{\partial A}{\partial K_{S}^{\mathrm{exp}}}\right)=\left(\begin{array}{ccc}
0 & 0 & 0\\
\frac{\partial A_{K_{S}e}}{\partial\mathrm{BR}(K_{S}e)} & 0 & \frac{\partial A_{K_{S}e}}{\partial\Gamma_{K_{S}}}\\
0 & 0 & 0\\
0 & 0 & 0\\
0 &\frac{\partial A_{K_{S}\mu}}{\partial\mathrm{BR}(K_{S}\mu)} & \frac{\partial A_{K_{S}\mu}}{\partial\Gamma_{K_{S}}}\\
0 & 0 & 0
\end{array}\right)~.
\end{equation}

\subsection{{\boldmath$K^{+}$} experimental inputs}

The $K^+$ decay width quoted in PDG is fitted from
Refs.~\cite{KLOE:2007wlh,Koptev:1995je,Ott:1971rs,Lobkowicz:1969mx,Fitch:1965zz}, and
the $K^+e$, $K^+\mu$ BRs from Refs.~\cite{KLOE:2007jte,Chiang:1972rp}. All of them, except two earlier experiments~\cite{Lobkowicz:1969mx,Chiang:1972rp}, were also used in the FlaviaNet analysis. The results read:
\begin{eqnarray}
	K_{+}^{\mathrm{exp}} & = & \left(\begin{array}{ccc}
		\mathrm{BR}(K^{+}e) & \mathrm{BR}(K^{+}\mu) & \Gamma_{K^{+}}\end{array}\right)^{T}\nonumber\\
	& = & \left(\begin{array}{ccc}
		5.07(4)\times10^{-2} & 3.352(33)\times10^{-2} & 5.3167(86)\times 10^{-14}\:\mathrm{MeV}\end{array}\right)^{T}
\end{eqnarray}
with the correlation matrix
\begin{equation}
\mathrm{Corr}(K_{+}^{\mathrm{exp}})=\left(\begin{array}{ccc}
1 & 0.8959847 & 0.01425396\\
& 1 & 0.01376368\\
&  & 1
\end{array}\right)~.
\end{equation}
The contribution of $K_{+}^{\mathrm{exp}}$ to the covariance matrix
of $A$ is given by:
\begin{equation}
\mathrm{Cov}(A)_{K_{+}^{\mathrm{exp}}}=\left(\frac{\partial A}{\partial K_{+}^{\mathrm{exp}}}\right)\cdot\mathrm{Cov}(K_{+}^{\mathrm{exp}})\cdot\left(\frac{\partial A}{\partial K_{+}^{\mathrm{exp}}}\right)^{T}
\end{equation}
where 
\begin{equation}
\left(\frac{\partial A}{\partial K_{+}^{\mathrm{exp}}}\right)=\left(\begin{array}{ccc}
0 & 0 & 0\\
0 & 0 & 0\\
\frac{\partial A_{K^{+}e}}{\partial\mathrm{BR}(K^{+}e)} & 0 & \frac{\partial A_{K^{+}e}}{\partial\Gamma_{K^{+}}}\\
0 & 0 & 0\\
0 & 0 & 0\\
0 & \frac{\partial A_{K^{+}\mu}}{\partial\mathrm{BR}(K^{+}\mu)} & \frac{\partial A_{K^{+}\mu}}{\partial\Gamma_{K^{+}}}
\end{array}\right).
\end{equation}

\subsection{Phase space factor}

The phase space factor is defined as\footnote{Notice that there is a typo in the $I_{K\ell}$
formula in many important references, e.g. Refs.~\cite{Antonelli:2010yf,Cirigliano:2011ny,CKM18,NewPS}.}:
\begin{equation}
I_{K\ell}=\int_{m_\ell^2}^{(M_K-M_\pi)^2}\frac{dt}{M_K^8}\bar{\lambda}^{3/2}\left(1+\frac{m_\ell^2}{2t}\right)
\left(1-\frac{m_\ell^2}{t}\right)^2\left[\bar{f}_+^2(t)+\frac{3m_\ell^2\Delta_{K\pi}^2}{(2t+m_\ell^2)
\bar{\lambda}}\bar{f}_0^2(t)\right]~,
\end{equation}
where $\bar{\lambda}=[t-(M_K+M_\pi)^2][t-(M_K-M_\pi)^2]$ and $\Delta_{K\pi}=M_K^2-M_\pi^2$. It probes
the $t$-dependence of the (rescaled) $K\pi$ vector and scalar form factors $\bar{f}_{+,0}(t)$,
which are obtained by fitting to the $K_{\ell 3}$ Dalitz plot. There are different ways to
parameterize the form factors, including the Taylor expansion~\cite{Antonelli:2010yf},
the $z$-parameterization~\cite{Hill:2006bq}, the pole parameterization~\cite{Lichard:1997ya} and
the dispersive parameterization~\cite{Bernard:2006gy,Bernard:2009zm,Abouzaid:2009ry}.
In this work, we take the results of the dispersive parameterization from the latest FlaviaNet updates which claim the
smallest uncertainty~\cite{CKM18,NewPS,PSCKM21}:
\begin{eqnarray}
	I_{K} & = & \left(\begin{array}{cccc}
		I_{K^0e} & I_{K^{+}e} & I_{K^0\mu} & I_{K^{+}\mu}\end{array}\right)^{T}\nonumber\\
	& = & \left(\begin{array}{cccc}
		0.15470(15)  & 0.15915(15) & 0.10247(15) & 0.10553(16)\end{array}\right)^{T}~.\nonumber\\
\end{eqnarray}
with the correlation matrix~\cite{PSCKM21}:
\begin{equation}
\mathrm{Corr}(I_{K})=\left(\begin{array}{cccc}
1 & 1 &0.530&0.521\\
  & 1 &0.530&0.521\\
  &   & 1  &1   \\
  &   &    &1
\end{array}\right)~.
\end{equation}
The contribution of $I_{K}$ to the covariance
matrix of $A$ is given by:
\begin{equation}
\mathrm{Cov}(A)_{I_{K}}=\left(\frac{\partial A}{\partial I_{K}}\right)\cdot\mathrm{Cov}(I_{K})\cdot\left(\frac{\partial A}{\partial I_{K}}\right)^{T}
\end{equation}
where 
\begin{equation}
\left(\frac{\partial A}{\partial I_{K}}\right)=\left(\begin{array}{cccc}
\frac{\partial A_{K_{L}e}}{\partial I_{K^{0}e}} & 0 & 0 & 0\\
\frac{\partial A_{K_{S}e}}{\partial I_{K^{0}e}} & 0 & 0 & 0\\
0 & \frac{\partial A_{K^{+}e}}{\partial I_{K^{+}e}} & 0 & 0\\
0 & 0 & \frac{\partial A_{K_{L}\mu}}{\partial I_{K^{0}\mu}} & 0\\
0 & 0 & \frac{\partial A_{K_{S}\mu}}{\partial I_{K^{0}\mu}} & 0\\
0 & 0 & 0 & \frac{\partial A_{K^{+}\mu}}{\partial I_{K^{+}\mu}}
\end{array}\right)~.
\end{equation}

\subsection{Isospin-breaking corrections}

The isospin-breaking correction $\delta_\mathrm{SU(2)}^{K\ell}$ is defined through the deviation of $f_+^{K\pi}(0)$ from $f_+^{K^0\pi^-}(0)$ (after scaling out the isospin factor $C_K$):
\begin{equation}
\delta_\mathrm{SU(2)}^{K\ell}=\left(\frac{f_+^{K\pi}(0)}{f_+^{K^0\pi^-}(0)}\right)^2-1~,
\end{equation}
so it resides in the $K^+$ channel only by construction. Upon neglecting small EM contributions,
it is given by~\cite{Antonelli:2010yf}:
\begin{equation}
  \delta_\mathrm{SU(2)}^{K^+\ell}=\frac{3}{2}\frac{1}{Q^2}\left[\frac{\hat{M}_K^2}{\hat{M}_\pi^2}
  +\frac{\chi_{p^4}}{2}\left(1+\frac{m_s}{\hat{m}}\right)\right]~,
\end{equation}
where $\hat{M}_{K,\pi}$ are the meson masses in the isospin limit, $Q^2=(m_s^2-\hat{m}^2)/(m_d^2-m_u^2)$,
and $\chi_{p^4}\simeq0.219$ is calculable in chiral perturbation theory (ChPT)~\cite{Gasser:1984ux}.
The pure Quantum Chromodynamics (QCD) mass parameters can only be obtained through lattice simulations.
Here we quote the most precise results of $Q$ and $m_s/\hat{m}$ from the latest
web-update~\cite{FLAGonline} of the Flavor Lattice Averaging Group (FLAG)
review~\cite{FlavourLatticeAveragingGroup:2019iem}:
\begin{eqnarray}
&&Q=23.3(5)~,\quad m_s/\hat{m}=27.42(12)~,\quad N_f=2+1~\qquad\quad\text{Refs.\cite{RBC:2014ntl,Durr:2010vn,Durr:2010aw,MILC:2009ltw,Fodor:2016bgu}}\nonumber\\
&&Q=24.0(8)~,\quad m_s/\hat{m}=27.23(10)~,\quad N_f=2+1+1~.\quad\text{Refs.\cite{Bazavov:2017lyh,EuropeanTwistedMass:2014osg,FermilabLattice:2014tsy,Giusti:2017dmp}}
\end{eqnarray}
Since $Q$ is by far the main contributor of the uncertainty in $\delta_\mathrm{SU(2)}^{K^+\ell}$, we choose the more precise data set from $N_f=2+1$ for numerical applications.
Meanwhile, Ref.~\cite{FlavourLatticeAveragingGroup:2019iem} did not provide the explicit values
of $\hat{M}_{K,\pi}$, so we quote them from the 2017 FLAG review~\cite{Aoki:2016frl}:
$\hat{M}_\pi=134.8(3)$~MeV, $\hat{M}_K=494.2(3)$~MeV. Putting pieces together, we have:
\begin{equation}
\delta_\mathrm{SU(2)}^{K^+\ell}=0.0457(20)~.
\end{equation}
The contribution of $\delta_{\mathrm{SU(2)}}$ to the covariance matrix
of $A$ is given by:
\begin{equation}
\mathrm{Cov}(A)_{\delta_{\mathrm{SU(2)}}}=\left(\frac{\partial A}{\partial\delta_{\mathrm{SU(2)}}}\right)\cdot\mathrm{Cov}(\delta_{\mathrm{SU(2)}})\cdot\left(\frac{\partial A}{\partial\delta_{\mathrm{SU(2)}}}\right)^{T}
\end{equation}
where 
\begin{equation}
\left(\frac{\partial A}{\partial\delta_{\mathrm{SU(2)}}}\right)=\left(\begin{array}{cccccc}
0 & 0 &
\frac{\partial A_{K^{+}e}}{\partial\delta_{\mathrm{SU(2)}}^{K^{+}\ell}} & 0 & 0 & \frac{\partial A_{K^{+}\mu}}{\partial\delta_{\mathrm{SU(2)}}^{K^{+}\ell}}
\end{array}\right)^T~.
\end{equation}

It should be pointed out that the parameter $Q$ can also be obtained phenomenologically. For instance, Ref.\cite{Colangelo:2018jxw} obtained $Q=22.1(7)$ from $\eta\rightarrow 3\pi$ decay, which is marginally discrepant with the FLAG average based on lattice calculations. We notice that different versions of FlaviaNet updates in the past few years had adopted different choices for their quark mass parameters: Ref.\cite{Moulson:2017ive} took the parameter $Q$ from lattice, while Refs.\cite{CKM18,NewPS} adopted the phenomenological value $Q=22.1(7)$ (together with a slightly different ChPT parameter $\chi_{p^4}\simeq 0.252$, of which the origin was not clearly explained), which returned a somewhat larger isospin breaking correction $\delta_\mathrm{SU(2)}^{K^+\ell}=0.0522(34)$.

\subsection{Long-distance EM corrections}

The last theory input is the long-distance EM correction. It was taken in the FlaviaNet review and its updates
from the ChPT calculation at $\mathcal{O}(e^2p^2)$~\cite{Cirigliano:2008wn}, with a theory
uncertainty of the order of $10^{-3}$. However, a novel framework based on Sirlin's representation
of RC~\cite{Sirlin:1977sv} was recently pioneered~\cite{Seng:2019lxf,Seng:2020jtz}. With this
framework and new lattice QCD inputs of the meson $\gamma W$-box diagrams~\cite{Feng:2020zdc,Ma:2021azh},
$\delta_\mathrm{EM}^{Ke}$ were re-evaluated with a significant increase in precision level
reaching $10^{-4}$~\cite{Seng:2021boy,Seng:2021wcf}. A similar update on $\delta_\mathrm{EM}^{K\mu}$ is
not yet available due to the more complicated error analysis but will be carried out in the
near future. Meanwhile, an important cross-check would be to compute the full $K_{\ell 3}$ RC (both the virtual and real corrections). This can be based on the existing technique that was proven successful in the study of the $K_{\mu 2}/\pi_{\mu 2}$ RCs~\cite{Giusti:2017dwk}, although its generalization to $K_{\ell 3}$ is expected to be more challenging and could take up to a decade to reach $10^{-3}$ precision~\cite{BoyleSnowmass}.

In this paper we choose to take the results of
$\delta_\mathrm{EM}^{K\mu}$ from ChPT, and that of $\delta_\mathrm{EM}^{Ke}$ from the new
calculation. Since these two evaluations are based on very different starting points, it is only
natural to assume that they are uncorrelated. 
In the $K\mu$ channels we have~\cite{Cirigliano:2008wn}:
\begin{eqnarray}
	\delta_{\mathrm{EM}}^{K\mu} & = & \left(\begin{array}{cc}
		\delta_{\mathrm{EM}}^{K^0\mu} & \delta_{\mathrm{EM}}^{K^{+}\mu}\end{array}\right)^{T}\nonumber\\
	& = & \left(\begin{array}{cc}
		1.40(22)\times10^{-2}& 0.016(250)\times10^{-2}\end{array}\right)^{T}
\end{eqnarray}
with the correlation matrix
\begin{equation}
\mathrm{Corr}(\delta_{\mathrm{EM}}^{K\mu})=\left(\begin{array}{cc}
1 &0.081\\
  & 1
\end{array}\right)~.
\end{equation}
Meanwhile, in the $Ke$ channels we have:
\begin{equation}
\delta_{\mathrm{EM}}^{Ke}=\left(\begin{array}{cc}
\delta_{\mathrm{EM}}^{K^0e} & \delta_{\mathrm{EM}}^{K^{+}e}\end{array}\right)^{T},
\end{equation}
where~\cite{Seng:2021boy,Seng:2021wcf}
\begin{eqnarray}
	\delta_{\mathrm{EM}}^{K^{0}e}& = & 1.16(2)_{\mathrm{sg}}(1)_{\mathrm{lat}}(1)_{\mathrm{NF}}(2)_{e^{2}p^{4}}\times10^{-2}\nonumber\\
	& = & 1.16(3)\times10^{-2}\nonumber\\
	\delta_{\mathrm{EM}}^{K^{+}e} & = & 0.21(2)_{\mathrm{sg}}(1)_{\left\langle r_{K}^{2}\right\rangle }(1)_{\mathrm{lat}}(4)_{\mathrm{NF}}(1)_{e^{2}p^{4}}\times10^{-2}\nonumber\\
	& = & 0.21(5)\times10^{-2}~.\label{eq:deltaEMKe}
\end{eqnarray}
The correlation matrix of $\delta_\mathrm{EM}^{K\ell}$ was not given in
Refs.~\cite{Seng:2021boy,Seng:2021wcf} and is derived here for the first time.

First of all, we realize that most of the uncertainties in Eq.~\eqref{eq:deltaEMKe}
were estimated through simple power-counting arguments on top of the
central values in each respective channel, so the most natural choice
is to take them as independent since assuming any correlation would
be equally arbitrary. However, there is a piece that has well-defined
correlations, namely the lattice calculation of the meson axial $\gamma W$-box diagrams,
which enters $\delta_\mathrm{EM}^{Ke}$ effectively through the low-energy constants (LECs)
in ChPT~\cite{Seng:2020jtz}: 
\begin{eqnarray}
  \delta_{\mathrm{EM}}^{K^{0}e} & = & 2e^{2}\left[\frac{4}{3}X_{1}-\frac{1}{2}\bar{X}_{6}^{\mathrm{phys}}
    \right]+...\nonumber\\
  \delta_{\mathrm{EM}}^{K^{+}e} & = & 2e^{2}\left[2\left(-\frac{2}{3}X_{1}-\frac{1}{2}
    \bar{X}_{6}^{\mathrm{phys}}\right)-\left(\frac{4}{3}X_{1}-\frac{1}{2}\bar{X}_{6}^{\mathrm{phys}}\right)
    \right]+... ~.
\end{eqnarray}
These two combinations of LECs were pinned down by two independent
lattice calculations of the axial $\gamma W$-box diagrams \cite{Feng:2020zdc,Ma:2021azh}:
\begin{eqnarray}
  \Box_{\gamma W}^{\pi} & : & -\frac{2}{3}X_{1}-\frac{1}{2}\bar{X}_{6}^{\mathrm{phys}}=-7.0(3)\times10^{-3}
  \nonumber\\
   \Box_{\gamma W}^{K} & : & \frac{4}{3}X_{1}-\frac{1}{2}\bar{X}_{6}^{\mathrm{phys}}=-11.3(5)\times10^{-3}~.
\end{eqnarray}
Therefore, the variations of $\delta_{\mathrm{EM}}^{Ke}$ due to the
lattice uncertainties are given by:
\begin{equation}
  \delta\left(\delta_{\mathrm{EM}}^{K^{0}e}\right)=2e^{2}\sigma_{K,\mathrm{lat}}~,\quad\delta
  \left(\delta_{\mathrm{EM}}^{K^{+}e}\right)=2e^{2}\left[2\sigma_{\pi,\mathrm{lat}}-\sigma_{K,\mathrm{lat}}\right]~,
\end{equation}
where $\sigma_{K,\mathrm{lat}}=0.5\times10^{-3}$ and  $\sigma_{\pi,\mathrm{lat}}=0.3\times10^{-3}$.
The two expressions above depend on a common quantity $\sigma_{K,\mathrm{lat}}$, which gives a
non-zero correlation:
\begin{equation}
  \left\langle \delta_{\mathrm{EM}}^{K^{0}e}\delta_{\mathrm{EM}}^{K^{+}e}\right\rangle -\left\langle
  \delta_{\mathrm{EM}}^{K^{0}e}\right\rangle \left\langle \delta_{\mathrm{EM}}^{K^{+}e}\right\rangle
  =-4e^{4}\sigma_{K,\mathrm{lat}}^{2}~.
\end{equation}
As a consequence, the correlation matrix reads:
\begin{equation}
\mathrm{Corr}(\delta_{\mathrm{EM}}^{Ke})=\left(\begin{array}{cc}
1 & \Delta\\
  & 1 
\end{array}\right)
\end{equation}
where
\begin{equation}
\Delta=-\frac{4e^{4}\sigma_{K,\mathrm{lat}}^{2}}{\sigma_{\delta_{\mathrm{EM}}^{K^{0}e}}\sigma_{\delta_{\mathrm{EM}}^{K^{+}e}}},
\end{equation}
with $\sigma_{\delta_{\mathrm{EM}}^{K^{0}e}}\simeq0.03\times10^{-2}$,
$\sigma_{\delta_{\mathrm{EM}}^{K^{+}e}}\simeq0.05\times10^{-2}$
as given above.

We may now combine all the independent long-distance EM corrections as:
\begin{equation}
\delta_{\mathrm{EM}}=\left(\begin{array}{cccc}
\delta_{\mathrm{EM}}^{K^0e} & \delta_{\mathrm{EM}}^{K^{+}e} & \delta_{\mathrm{EM}}^{K^0\mu}
& \delta_{\mathrm{EM}}^{K^{+}\mu}\end{array}\right)^{T}.
\end{equation}
Its correlation matrix is given by the following block-diagonal matrix:
\begin{equation}
\mathrm{Corr}(\delta_{\mathrm{EM}})=\left(\begin{array}{cc}
\mathrm{Corr}(\delta_{\mathrm{EM}}^{Ke}) & 0\\
0 & \mathrm{Corr}(\delta_{\mathrm{EM}}^{K\mu})
\end{array}\right).
\end{equation}
The contribution of $\delta_{\mathrm{EM}}$ to the covariance matrix
of $A$ is given by:
\begin{equation}
\mathrm{Cov}(A)_{\delta_{\mathrm{EM}}}=\left(\frac{\partial A}{\partial\delta_{\mathrm{EM}}}\right)\cdot\mathrm{Cov}(\delta_{\mathrm{EM}})\cdot\left(\frac{\partial A}{\partial\delta_{\mathrm{EM}}}\right)^{T}
\end{equation}
where 
\begin{equation}
\left(\frac{\partial A}{\partial\delta_{\mathrm{EM}}}\right)=\left(\begin{array}{cccc}
\frac{\partial A_{K_{L}e}}{\partial\delta_{\mathrm{EM}}^{K^{0}e}} & 0 & 0 & 0\\
\frac{\partial A_{K_{S}e}}{\partial\delta_{\mathrm{EM}}^{K^{0}e}} & 0 & 0 & 0\\
0 & \frac{\partial A_{K^{+}e}}{\partial\delta_{\mathrm{EM}}^{K^{+}e}} & 0 & 0\\
0 & 0 & \frac{\partial A_{K_{L}\mu}}{\partial\delta_{\mathrm{EM}}^{K^{0}\mu}} & 0\\
0 & 0 & \frac{\partial A_{K_{S}\mu}}{\partial\delta_{\mathrm{EM}}^{K^{0}\mu}} & 0\\
0 & 0 & 0 & \frac{\partial A_{K^{+}\mu}}{\partial\delta_{\mathrm{EM}}^{K^{+}\mu}}
\end{array}\right)~.
\end{equation}

\subsection{Final result}

\begin{table}
	\begin{centering}
		\begin{tabular}{|c|c|cccccc|}
			\hline 
			& $|V_{us}f_{+}^{K}(0)|$ & \multicolumn{6}{c|}{Correlation Matrix}\tabularnewline
			\hline 
			\hline 
			$K_{L}e$ & $0.21617(46)_{\mathrm{exp}}(10)_{I_{K}}(3)_{\delta_{\mathrm{EM}}}\,\,\,\,$ & 1 & 0.021 & 0.025 & 0.519 & 0.004 & 0.017\tabularnewline
			\cline{1-2} 
			$K_{S}e$ & $0.21530(122)_{\mathrm{exp}}(10)_{I_{K}}(3)_{\delta_{\mathrm{EM}}}\,\,$ &  & 1 & 0.009 & 0.012 & 0.000 & 0.006\tabularnewline
			\cline{1-2} 
			$K^{+}e$ & $\,\,\,\,\,\,\,\,\,\,\,\,\,\,\,\,\,\,0.21714(88)_{\mathrm{exp}}(10)_{I_{K}}(21)_{\delta_{\mathrm{SU(2)}}}(5)_{\delta_{\mathrm{EM}}}$ &  &  & 1 & 0.016 & 0.002 & 0.871\tabularnewline
			\cline{1-2} 
			$K_{L}\mu$ & $0.21664(50)_{\mathrm{exp}}(16)_{I_{K}}(24)_{\delta_{\mathrm{EM}}}\,\,$ &  &  &  & 1 & 0.029 & 0.047\tabularnewline
			\cline{1-2} 
			$K_{S}\mu$ & $\,\,0.21265(466)_{\mathrm{exp}}(16)_{I_{K}}(23)_{\delta_{\mathrm{EM}}}\,$ &  &  &  &  & 1 & 0.006\tabularnewline
			\cline{1-2} 
			$K^{+}\mu$ & $\,\,\,\,\,\,\,\,\,\,\,\,\,\,\,\,\,\,\,\,\,\,\,\,\,0.21703(108)_{\mathrm{exp}}(16)_{I_{K}}(21)_{\delta_{\mathrm{SU(2)}}}(26)_{\delta_{\mathrm{EM}}}$ &  &  &  &  &  & 1\tabularnewline
			\hline 
			Average: $Ke$ & $0.21626(40)_K(3)_\mathrm{HO}$ &  &  &  &  &  & \multicolumn{1}{c}{}\tabularnewline
			\cline{1-2} 
			Average: $K\mu$ & $0.21667(52)_K(3)_\mathrm{HO}$ &  &  &  &  &  & \multicolumn{1}{c}{}\tabularnewline
			\cline{1-2} 
			Average: tot & $0.21635(39)_{K}(3)_{\mathrm{HO}}$ &  &  &  &  &  & \multicolumn{1}{c}{}\tabularnewline
			\cline{1-2} 
		\end{tabular}
		\par\end{centering}
	        \caption{\label{tab:Vusf} Individual values and weighted average of $|V_{us}f_+^K(0)|$,                 with independent uncertainties displayed separately.  Notice that the uncertainties
                  from the BR and $\Gamma_K$ are generally correlated, so we display only their
                  combined uncertainty as ``exp''.}
	
\end{table}

With the above, we may obtain the values of $|V_{us}f_+^K(0)|$ from each channel, which are
summarized in the left panel of Table~\ref{tab:Vusf}.
Making use of the total covariance matrix of $A$:
\begin{equation}
  \mathrm{Cov}(A)=\mathrm{Cov}(A)_{K_{L}^{\mathrm{exp}}}+\mathrm{Cov}(A)_{K_{S}^{\mathrm{exp}}}
  +\mathrm{Cov}(A)_{K_{+}^{\mathrm{exp}}}+\mathrm{Cov}(A)_{I_{K}}+\mathrm{Cov}(A)_{\delta_{\mathrm{SU(2)}}}
  +\mathrm{Cov}(A)_{\delta_{\mathrm{EM}}}~,
\end{equation}
from which the total correlation matrix in the right panel of Table~\ref{tab:Vusf} can be calculated,
the weighted average of $|V_{us}f_+^K(0)|$ can then be obtained using Eq.\eqref{eq:weightedmean}. Given the different theory statuses of $K_{e3}$ and $K_{\mu 3}$, we present simultaneously the average values of $|V_{us}f_+^K(0)|$ by weighting over the $Ke$ channels, the $K\mu$ channels, and both.
Notice that the uncertainty from $S_{\mathrm{EW}}$ is common to all channels and does not enter
the weighting process. Therefore, we choose to display it only in the weighted averages,
i.e. the last three rows in Table~\ref{tab:Vusf}. We find that the $Ke$ and $K\mu$ averages agree well with each other within uncertainties.
Finally, the 2020 website update of the FLAG review quoted~\cite{FLAGonline}:
\begin{equation}
\begin{array}{ccc}
|f_+^K(0)|=0.9698(17) & N_f=2+1+1 & \quad\quad\text{Refs.\cite{Carrasco:2016kpy,Bazavov:2018kjg}}\\
|f_+^K(0)|=0.9677(27) & N_f=2+1 & \quad\quad\text{Refs.\cite{Bazavov:2012cd,Boyle:2015hfa}}\\
|f_+^K(0)|=0.9560(57)(62) & N_f=2 & \text{Ref.\cite{Lubicz:2009ht}}
\end{array}
\end{equation}
We choose the most precise value from $N_f=2+1+1$ for numerical applications.
With that we obtain:
\begin{equation}
|V_{us}|_{K_{\ell 3}}=0.22309(40)_K(39)_\mathrm{lat}(3)_\mathrm{HO}~.\label{eq:Vus}
\end{equation}

Let us discuss the results above. First, both the central value and the total uncertainty in
the weighted average of $|V_{us}f_+^K(0)|$ experience no significant change compared to those
in previous reviews (e.g. $0.21654(41)$ in Ref.~\cite{Moulson:2017ive}), but not the
composition of uncertainties in each channel. In our latest analysis, the combined experimental
uncertainty from the kaon lifetime and BRs are by far the dominant source of uncertainty
in all channels. This is quite different from a few years ago, wherein some channels (e.g. $K^+e$)
the theory and experimental uncertainties are comparable.
Such changes are mainly due to the improved theory precision in $\delta_\mathrm{SU(2)}^{K\ell}$
and $\delta_\mathrm{EM}^{Ke}$.  

We may also review the status of the top-row CKM unitarity. The best extraction of $|V_{ud}|$
comes from superallowed $0^+\to 0^+$ beta decays, but its precise value depends on the theory
inputs of the single-nucleon RC and nuclear structure corrections. In particular, it was recently
pointed that several potentially large new nuclear corrections (NNCs) that reside in the
nuclear $\gamma W$-box diagrams were missed in the existing nuclear structure
calculations~\cite{Seng:2018qru,Gorchtein:2018fxl}; their true sizes are poorly understood
and are at present only roughly estimated based on a simple Fermi gas model. After
alerting the readers about the possible (small) quantitative difference due to different
choices of theory inputs, let us quote, just for this work, the result from the latest
review by Hardy and Towner~\cite{Hardy:2020qwl}:
\begin{equation}
|V_{ud}|_{0^+}=0.97373(11)_\mathrm{exp}(9)_\mathrm{RC}(27)_\mathrm{NS}~,\label{eq:Vud}
\end{equation}
where ``exp'' is the combined uncertainty from experiment and the so-called ``outer''
correction, ``RC'' the theory uncertainty from the single-nucleon (inner) RC, and
``NS'' the nuclear structure uncertainty that originates primarily from the NNCs.
Combining Eqs.\eqref{eq:Vus} and \eqref{eq:Vud} gives:
\begin{equation}
|V_{ud}|_{0^+}^2+|V_{us}|_{K_{\ell 3}}^2-1=-0.0021(2)_{V_{ud},\mathrm{exp}}(2)_{V_{ud},\mathrm{RC}}(5)_{V_{ud},\mathrm{NS}}(2)_{V_{us},K}(2)_{V_{us},\mathrm{lat}}~,
\end{equation}
while the SM prediction (after neglecting the small $|V_{ub}|^2$) is 0. The above indicates an apparent anomaly in the top-row CKM unitarity with the significance level of $3.2\sigma$, which could increase to as much as $5.6\sigma$ if we imagine that the NS uncertainty was significantly reduced while the central value of $|V_{ud}|$ remained unchanged. This provides a strong motivation for nuclear theorists to perform \textit{ab-initio} calculations of the NS correction in superallowed beta decays to reduce its theory uncertainty.

\section{ $|V_{us}f_{+}^{K}(0)|/|V_{ud}f_{+}^{\pi}(0)|$ and $|V_{us}/V_{ud}|$ from $K/\pi$ semileptonic decays}
\label{sec:Vus/Vud}

In addition to its contribution to an apparent violation of the top-row CKM unitarity, Eq.~\eqref{eq:Vus} also
shows a direct disagreement at the level $\sim 2.8\sigma$ with the same quantity extracted from
$K_{\mu 2}$ decay: $|V_{us}|_{K_{\mu 2}}=0.2252(5)$~\cite{Zyla:2020zbs}. The latter is
obtained from the ratio $R_A=\Gamma(K_{\mu 2})/\Gamma(\pi_{\mu 2})$, which gives the value
of $|V_{us}f_{K^+}|/|V_{ud}f_{\pi^+}|$, with $f_{K^+}$ and $f_{\pi^+}$ the decay constants of the
charged kaon and pion~\cite{Marciano:2004uf}:
\begin{equation}
\left|\frac{V_{us}f_{K^+}}{V_{ud}f_{\pi^+}}\right|=0.23871(20)_\mathrm{RC}\times R_A^{1/2}~,\label{eq:RA}
\end{equation}
where the theory uncertainty of $0.084\%$ at the right-hand side originates from residual
long-distance RCs that do not cancel in the ratio. This residual RC has been calculated using both ChPT~\cite{Cirigliano:2011tm} and lattice QCD~\cite{Giusti:2017dwk} with comparable sizes of theory uncertainties. The two calculations agree well with each other, showing that the theory error in this input is under good control. Following PDG, we utilize the ChPT input in Eq.\eqref{eq:RA} for illustration (throughout this work we add nothing new in the $R_A$ analysis; everything is the same as in the 2021 online version of the PDG review).

The above discrepancy may indicate the presence of BSM effects or possible unidentified SM
corrections that are not reflected in the existing error estimation, such as a smaller value for $|f_+^K(0)|$ outside the range of the quoted $N_f=2+1+1$ lattice QCD result. To further explore these
possibilities, in particular the latter, Ref.~\cite{Czarnecki:2019iwz} suggested to study a new
ratio $R_V=\Gamma(K_{\ell 3})/\Gamma(\pi_{e3})$ (which takes different values in different
$K\ell$ channels), where the $V$ denotes the fact that such decays are due to weak vector current interactions. Like $R_A$, it results from a ratio of weak interaction meson decays (induced by vector rather than axial-vector
interactions) for which theoretical uncertainties partially cancel. A comparison of $R_V$ and $R_A$ can, in principle, unveil the influence of BSM physics. 

We first recall the SM prediction of the $\pi_{e3}$ decay width:
\begin{equation}
  \Gamma_{\pi_{e3}}=\frac{G_F^2|V_{ud}|^2M_{\pi^+}^5|f_+^\pi(0)|^2}{64\pi^3}(1+\mathrm{RC}_\pi)I_\pi~.
  \label{eq:pie3width}
\end{equation}
The left-hand side is calculated from the experimental measurement of the charged pion lifetime
and the semileptonic decay BR~\cite{Zyla:2020zbs}:
\begin{equation}
\tau_{\pi^{+}}  =  2.6033(5)\times10^{-8}\:\mathrm{s}~,\quad 
\mathrm{BR}(\pi_{e3})  =  1.038(6)\times10^{-8}~.
\end{equation}
Notice the slight modification of the BR from the PDG value that took into account the effect
of the updated $\mathrm{BR}(\pi_{e2})$ normalization~\cite{Czarnecki:2019iwz}.
On the right-hand side, $|f_+^\pi(0)|$ is the $\pi^+\to \pi^0$ form factor at zero momentum
transfer which equals 1 in the isospin limit (isospin-breaking correction is negligible due to the
Behrends-Sirlin-Ademollo-Gatto theorem~\cite{Behrends:1960nf,Ademollo:1964sr}).  In this limit, the phase space integral $I_\pi$ is calculable:
\begin{eqnarray}
I_{\pi}&=&\int_{2\sqrt{r_\pi}}^{1+r_\pi-r_e} dz\int_{c(z)-d(z)}^{c(z)+d(z)}dy\left[4(1-y)(y+z-1)+r_e(4y+3z-3)-4r_\pi+r_e(r_\pi-r_e)\right]\nonumber\\
 &=& 7.3764\times10^{-8}~, 
\end{eqnarray}
where
\begin{equation}
r_\pi=\frac{M_{\pi^0}^2}{M_{\pi^+}^2}~,\:\: r_e=\frac{m_e^2}{M_{\pi^+}^2}~,\:\:c(z)=\frac{(2-z)(1+r_e+r_\pi-z)}{2(1+r_\pi-z)}~,\:\: d(z)=\frac{\sqrt{z^2-4r_\pi}(1+r_\pi-r_e-z)}{2(1+r_\pi-z)}~,
\end{equation}
in analogy to the well-known $K_{\ell 3}$ phase space formula (see, e.g. Ref.\cite{Seng:2021wcf}). $\mathrm{RC}_\pi$ is
the electroweak RC in the pion semileptonic decay which was recently determined to high
precision with lattice QCD: $\mathrm{RC}_\pi=0.0332(1)_{\gamma W}(3)_\mathrm{HO}$~\cite{Feng:2020zdc}.
Combining Eqs.~\eqref{eq:Kl3master} and \eqref{eq:pie3width} gives:
\begin{equation}
  \left|\frac{V_{us}f_{+}^{K}(0)}{V_{ud}f_{+}^{\pi}(0)}\right|_{K\ell}=\sqrt{\frac{3}{C_{K}^{2}}
    \left(\frac{M_{\pi^{+}}}{M_{K}}\right)^{5}\frac{I_{\pi}}{I_{K\ell}}\frac{1
      +\mathrm{RC_{\pi}}}{S_{\mathrm{EW}}(1+\delta_{\mathrm{EM}}^{K\ell}+\delta_{\mathrm{SU(2)}}^{K\ell})}}
  \times (R_V^{K\ell})^{1/2}~,\label{eq:RV}
\end{equation}
which provides a measure of $|V_{us} f_+^K(0)|/|V_{ud}f_+^\pi(0)|$. 

There are several benefits in studying $R_V$. First, uncertainties from short-distance electroweak
RCs (contained in $\mathrm{RC}_\pi$ and $S_\mathrm{EW}$, although they are numerically smaller than the other SM theory uncertainties, e.g. those coming from $I_K$ and $\delta_\mathrm{SU(2)}$) as well as BSM effects that are
common to the numerator and denominator (e.g. those correcting $G_F$ through the muon lifetime)
cancel each other in the $R_V$ ratio. This means as follows: should one observes a significant discrepancy
between the values of $|V_{us}/V_{ud}|$ obtained from $R_V$ and $R_A$, then its possible BSM
explanations would be more limited than those which could be used to explain the discrepancy
between the values of $|V_{us}|$ from $K_{\mu 3}$ and $R_A$. This makes $R_V$ a useful gauge to
search for possibly large non-universal systematic effects, especially those from the SM.
Second, the recent improvements in SM theory precision, in particular the electroweak RCs,
makes $R_V$ an extremely clean observable from the theory aspect. Consider, for instance,
the $K_Le$ channel. Substituting all the SM theory inputs we discussed above, one obtains:
\begin{equation}
\left|\frac{V_{us}f_+^K(0)}{V_{ud}f_+^\pi(0)}\right|_{K_Le}=4.9786(24)_{I_K}(7)_{\mathrm{RC}_K}(2)_{\mathrm{RC}_\pi}\cdot 10^{-5}\times (R_V^{K_Le})^{1/2}~.
\end{equation} 
We see that the total theory uncertainty on the right-hand side is only 0.051\%, which is
already better than $R_A$. Moreover, the above is only for one channel in $K_{\ell 3}$.
A further reduction of uncertainty is achieved once all six channels are weighted over.

Upon substituting the experimental inputs into Eqs.\eqref{eq:RA} and \eqref{eq:RV}, we find:
\begin{eqnarray}
\left|\frac{V_{us}f_{K^+}}{V_{ud}f_{\pi^+}}\right|&=&0.27600(29)_\mathrm{exp}(23)_\mathrm{RC}~,\quad \text{0.13\% precision} \nonumber\\
\left|\frac{V_{us}f_+^K(0)}{V_{ud}f_+^\pi(0)}\right|&=&0.22216(64)_{\mathrm{BR}(\pi_{e3})}(40)_K(2)_{\tau_{\pi^+}}(1)_{\mathrm{RC}_\pi}~,\quad \text{0.34\% precision}
\label{eq:RAvsRV}
\end{eqnarray}
where the first line is from $R_A$, and the second line is from $R_V$ weighted over all six channels. The precision of the latter is limited primarily by the uncertainty in $\mathrm{BR}(\pi_{e3})$,
and secondarily by the $K_{\ell 3}$ experiments. Future improvements of the experimental precision
in these areas are therefore urgently needed.

At this point it is interesting to discuss the relevance with respect to PIONEER, the proposed next-generation
experiment for rare pion decays which may take place in PSI or TRIUMF~\cite{ArevaloSnowmass,Pioneer}. It is originally designed for an
improved measurement of the ratio $R_{e/\mu}=\Gamma(\pi^+\to e^+\nu(\gamma))/\Gamma(\pi^+\to\mu^+\nu(\gamma))$
to test  lepton universality, but the optimized detector is also ideal for a high-precision measurement
of $\mathrm{BR}(\pi_{e3})$. The current best measurement of the latter is from the
PIBETA experiment~\cite{Pocanic:2003pf}, which leads to the following extracted value of
$|V_{ud}|$~\cite{Feng:2020zdc}:
\begin{equation}
|V_{ud}|_{\pi_{e3}}=0.9740(28)_\mathrm{exp}(1)_\mathrm{th}~.
\end{equation}
Despite being theoretically clean, it is 10 times less precise than the superallowed beta decay extraction (see Eq.\eqref{eq:Vud}). To make $|V_{ud}|_{\pi_{e3}}$ competitive requires a 10 times reduction of the experimental uncertainty, which may require 100 times the statistics of the existing measurement and comparable reduction of systematics and backgrounds, a very ambitious long-term goal. However, the introduction of $R_V$ provides a new physical significance to $\pi_{e3}$, not just in terms of $|V_{ud}|$ but also $|V_{us}/V_{ud}|$. With this, it is most beneficial to plan the next-generation $\pi_{e3}$ experiment for two stages:
\begin{itemize}
	\item The first stage would primarily aim to improve the precision of the ratio, $R_{e/\mu}$ (its primary goal) by an order of magnitude.  That same phase could be used to improve the precision of
	BR$(\pi_{e3})$ by a factor of 3 or better compared to the existing PIBETA result. That would reduce the uncertainty in $R_V$ to a level comparable to $R_A$, making for an interesting confrontation. If accompanied by future improvement in $K_{\ell 3}$ experiments, $R_V$ could eventually surpass $R_A$ as the primary means to constrain  $|V_{us}/V_{ud}|$.
	\item In the second stage an overall improvement of a factor of 10 improvement in the $\mathrm{BR}(\pi_{e3})$ precision is required to compete with superallowed beta decays for precision in extracting $V_{ud}$.
	It is, however, much more challenging and is not yet at the achievable level in the present technical design~\cite{Hertzog}.
\end{itemize}

\begin{figure}
	\begin{centering}
		\includegraphics[scale=0.8]{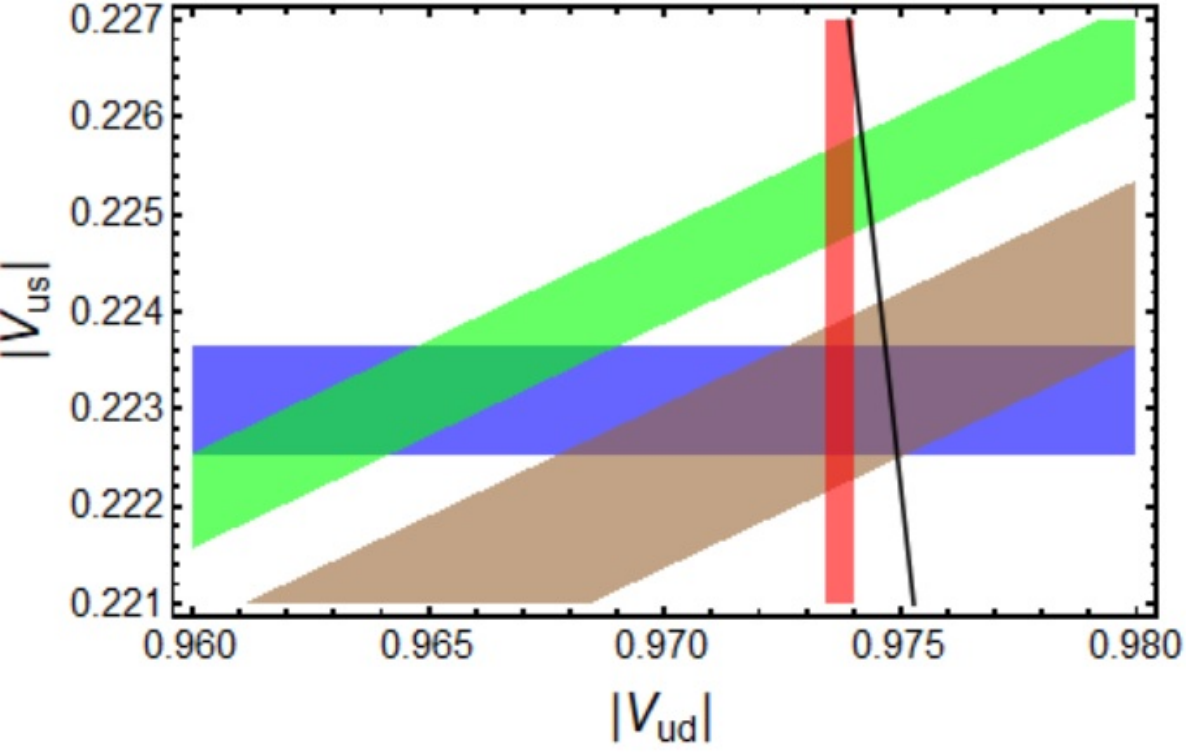}
		\hfill
		\par\end{centering}
	\caption{\label{fig:VudVus}A combined plot of $|V_{ud}|$ from superallowed beta decays (red band), $|V_{us}|$ from $K_{\ell 3}$ (blue band), $|V_{us}/V_{ud}|$ from $R_A$ (green band) and $R_V$ (brown band), together with the first-row CKM unitarity requirement (black line). Notice that the blue and brown bands are highly correlated as they rely on the same set of inputs from $K_{\ell 3}$, so they should not be taken as independent constraints.  }
\end{figure}

We close this section by reporting the current extracted values of $|V_{us}/V_{ud}|$
from $R_A$ and $R_V$ respectively, by supplementing Eq.~\eqref{eq:RAvsRV} with relevant
lattice QCD inputs~\cite{FLAGonline}:
\begin{equation}
\begin{array}{ccc}
|f_{K^+}/f_{\pi^+}|=1.1932(21) & \quad N_f=2+1+1 & \quad\text{Refs.\cite{Bazavov:2017lyh,Dowdall:2013rya,Carrasco:2014poa,Miller:2020xhy}}\\
|f_+^K(0)/f_+^\pi(0)|\approx |f_+^K(0)|=0.9698(17)~. & \quad N_f=2+1+1 & \text{Refs.\cite{Carrasco:2016kpy,Bazavov:2018kjg}}
\end{array}
\end{equation} 
They give:
\begin{equation}
|V_{us}/V_{ud}|=\left\{ \begin{array}{ccc}
0.23131(41)_\mathrm{lat}(24)_\mathrm{exp}(19)_\mathrm{RC} &  & \text{from $R_A$}\\
0.22908(66)_{\mathrm{BR}(\pi_{e3})}(41)_K(40)_\mathrm{lat}(2)_{\tau_{\pi^+}}(1)_{\mathrm{RC}_\pi} &  & \text{from $R_V$}
\end{array}\right.~.
\end{equation}
The difference between the two determinations is at the level of $2.2\sigma$. All the determinations of $|V_{us}|$, $|V_{ud}|$ and their ratio quoted in this paper are summarized in Fig.\ref{fig:VudVus}, from which the mutual disagreements between different determinations and the deviations from the first-row CKM unitary requirement are clearly shown.

\section{Conclusions}
\label{sec:con}

This work updates the values of $|V_{us}|$ and $|V_{us}/V_{ud}|$ determined from kaon and pion
semileptonic decays using the most recent inputs from theory and experiment. The uncertainties in these quantities have been experiment- and lattice-dominated, which is even more the case in the recent years due to the more precise SM electroweak theory inputs. Their values along with $V_{ud}$ from superallowed beta decays correspond to 2--3$\sigma$ deviations from CKM unitary and related axial current induced weak decays. Those differences may provide hints of BSM physics or deficiencies in SM theory or experiment.  Such anomalies provide a strong
motivation for future improvements of the experimental precision of the $\pi_{e3}$ BR
as well as the kaon lifetimes and $K_{\ell 3}$ BRs. 

The experiment-dominated uncertainties do not imply that future improvements from the theory
side are not important. It is quite the opposite; the aforementioned anomalies require
us to carefully reexamine all the SM theory inputs in order to ensure that no unexpected
large systematic errors exist. This was recently done for the long-distance EM corrections to the
$K_{e3}$ decay rate and no large corrections were found. Other inputs, such as the lattice
calculations of $f_+^K(0)$ and $f_{K^+}/f_{\pi^+}$, should be cross-checked with the same level of rigor.
There are several other theory works that remain to be done for completeness and
internal consistency: For instance, the reevaluation of the EM corrections should be generalized
to the $K\mu$ channels, and the fitting of the $K\pi$ form factors should, in principle, also be
updated to account for the modified EM corrections to the $K_{\ell 3}$ Dalitz plot.

\begin{acknowledgments}

This work is supported in
part by the Deutsche Forschungsgemeinschaft (DFG, German Research
Foundation) and the NSFC through the funds provided to the Sino-German Collaborative Research
Center TRR110 “Symmetries and the Emergence of Structure in QCD” (DFG Project-ID 196253076 -
TRR 110, NSFC Grant No. 12070131001) (UGM and CYS), by the Alexander von Humboldt Foundation
through the Humboldt Research Fellowship (CYS), by the Chinese Academy of Sciences
(CAS) through a President's International Fellowship Initiative (PIFI) (Grant No. 2018DM0034)
and by the VolkswagenStiftung (Grant No. 93562) (UGM), by EU Horizon 2020 research and
innovation programme, STRONG-2020 project under grant agreement No 824093 (UGM), and by the
U.S. Department of Energy under Grant DE-SC0012704 (WJM). 
  
\end{acknowledgments}

\begin{appendix}
	
\section{\label{sec:tools}Mathematical Tools}

In this Appendix, we review all the mathematical tools needed in
this work.

\subsection{Covariance matrix and correlation matrix}

For a set of variables $X=(x_{1},x_{2},...,x_{n})^{T}$, we define
the symmetric covariance matrix $\mathrm{Cov}(X)$ as:
\begin{equation}
\mathrm{Cov}(X)_{ij}=\left\langle x_{i}x_{j}\right\rangle -\left\langle x_{i}\right\rangle \left\langle x_{j}\right\rangle~.
\end{equation}
In particular, the diagonal terms give the variance of $x_{i}$: 
\begin{equation}
\sigma_{i}^{2}=\mathrm{Cov}(X)_{ii}~.
\end{equation}
We also define the symmetric correlation matrix $\mathrm{Corr}(X)$ as:
\begin{equation}
\mathrm{Corr}(X)_{ij}=\frac{\left\langle x_{i}x_{j}\right\rangle -\left\langle x_{i}\right\rangle \left\langle x_{j}\right\rangle }{\sigma_{i}\sigma_{j}}~.
\end{equation}
Its diagonal elements are always 1, while the off-diagonal
elements range between $-1$ and 1. Its relation to the covariance matrix
is given by:
\begin{equation}
\mathrm{Cov}(X)=\sigma_{X}\cdot\mathrm{Corr}(X)\cdot\sigma_{X}~,\label{eq:CorrtoCov}
\end{equation}
where $\sigma_{X}\equiv\mathrm{diag}(\sigma_{1},\sigma_{2},...,\sigma_{n})$. 

\subsection{Propagation of the covariance matrix}

For a set of variables $F=(f_{1},f_{2},...,f_{m})^{T}$ that are functions
of $X$ (i.e. $f_{i}=f_{i}(x_{1},...,x_{n})$), given the covariance
matrix of $X$ we can immediately obtain the covariance matrix of
$F$ as:
\begin{equation}
  \mathrm{Cov}(F)=\left(\frac{\partial F}{\partial X}\right)\cdot\mathrm{Cov}(X)\cdot
  \left(\frac{\partial F}{\partial X}\right)^{T}
\end{equation}
where $\partial F/\partial X$ is a $m\times n$ matrix, with matrix elements:
\begin{equation}
\left(\frac{\partial F}{\partial X}\right)_{ij}=\frac{\partial f_{i}}{\partial x_{j}}~.
\end{equation}
If $F=(f_{1},f_{2},...,f_{m})^{T}$ depends on two {\em independent} sets
of variables $X=(x_{1},x_{2},...,x_{n})^{T}$ and $Y=(y_{1},y_{2},...,y_{n'})^{T}$
with their respective covariance matrices given, then the covariance
matrix of $F$ is simply the sum of the two contributions:
\begin{equation}
  \mathrm{Cov}(F)=\left(\frac{\partial F}{\partial X}\right)\cdot\mathrm{Cov}(X)\cdot
  \left(\frac{\partial F}{\partial X}\right)^{T}+\left(\frac{\partial F}{\partial Y}\right)
  \cdot\mathrm{Cov}(Y)\cdot\left(\frac{\partial F}{\partial Y}\right)^{T}~.
\end{equation}
This is also generalizable if $F$ is a function of more than two independent
sets of variables.

For definiteness, throughout this work, we always calculate partial derivatives
numerically as:
\begin{equation}
\frac{\partial f}{\partial a}\approx\frac{f(a+\delta a)-f(a-\delta a)}{2\delta a}~.
\end{equation}

\subsection{Weighted average }

If $X=(x_{1},x_{2},...,x_{n})^{T}$ has a covariance matrix $\mathrm{Cov}(X)$,
then the weighted average between $x_{1},...,x_{n}$ is given by:
\begin{equation}
\bar{x}=\sigma_{\bar{x}}^{2}\left(J^{T}WX\right)\label{eq:weightedmean}
\end{equation}
where the variance of $\bar{x}$ is given by:
\begin{equation}
\sigma_{\bar{x}}^{2}=\left(J^{T}WJ\right)^{-1}~.
\end{equation}
Here we have defined $W=\left[\mathrm{Cov}(X)\right]^{-1}$ and $J=(1,...,1)^{T}$
(length = $n$). 	
	
\end{appendix}

\end{document}